\documentstyle[aps,prl,twocolumn]{revtex}

\begin{document}


\draft
\vspace{0.5in}
\title{Approximate boost of the relativistic two-body problem in instant formalism}


\author{S. J. Wallace }
\address{Department of Physics and Center for Theoretical Physics,
University of Maryland, College Park, MD 20742}

\date{\today}
\maketitle

\begin{abstract}
Considering two spinless particles, a simple, approximate boost rule is derived 
that is sufficient to keep the mass invariant and 
to relate interactions, vertex functions, wave functions and t-matrices of
the instant two-body problem in an arbitrary frame to those in the c.m. frame.
The boost generator ${\bf K}$ that is used provides a solution to
the Poincar\'{e} algebra except for the commutator between two components
of ${\bf K}$, for which it provides an approximate solution.     
\end{abstract}




The two-body boost problem in the instant form of relativistic quantum mechanics
is to constrain the interaction so that the Poincar\'{e} generators
satisy the commutation rules that are required for Poincar\'{e} invariance.
The central issue in boosting an instant form of dynamics was
identified by Dirac~\cite{Dirac49}: the generator
of boosts, ${\bf K}$, and the hamiltonian, $H$, both must contain the interaction, $v$.
Such a boost is called dynamical and
the commutation rule between the dynamical boost generator and hamiltonian
involves $v^2$ terms.

Bakamjian and Thomas~\cite{Bakamjian53} showed that the complete commutator algebra 
for two particles could be satisfied
by introducing the interaction in a mass 
operator, $h = h_0 + v_c$, which is the same as the hamiltonian in the c.m. frame.
The Bakamjian-Thomas hamiltonian in an arbitrary frame is  
$H = \sqrt{h^2 + {\bf P}^2}$, 
where ${\bf P}$ is the total momentum.   From this one 
defines the interaction 
in an arbitrary frame as $v_{BT} \equiv H - H_0$, 
where $H_0$ is the hamiltonian
for two free particles, $H_0 = \epsilon_1 + \epsilon_2$. 
Consequently there is a 
nonlinear relation between the interaction in the c.m. frame 
and the interaction and kinetic energy terms in another frame.  

Foldy and collaborators~\cite{Foldy61,Kracjik74} showed that 
a linear boost relation for the interaction in any frame 
could be obtained by working
to leading order in an expansion in
powers of the ratio of momenta to the mass of a particle.
See also 
Friar.~\cite{Friar73,Friar80}  Coester, Pieper and Serduke \cite{CPS75}
also proposed an approximate boost rule that is linear in $v$.

An alternative that avoids 
constructing the interaction in other than the c.m. frame is to
diagonalize the mass operator and use its eigenvalues in the construction
of an irreducible representation of the Poincar\'{e} group for each
mass eigenvalue.  This latter point of view is stressed in the
review of relativistic hamiltonian dynamics by Keister and
Polyzou.~\cite{Keister91}  It produces a kinematical form of
boost but of course the mass eigenvalue
depends upon the interaction.  

In quantum field theory, the boost of a mass eigenstate such as a
Bethe-Salpeter vertex function also is kinematical.  There is
a linear relation between the interactions in different frames that
depends upon the Lorentz transformation of momenta and spins. 
Also in this case the boost implicity depends upon the interaction through
the mass eigenvalue, $M$, which enters the boost velocity
$\beta = {\bf P}/E_{{\bf P}}$ via 
$E_{{\bf P}} = \sqrt{M^2 + {\bf P}^2}$.

If the Bethe-Salpeter formalism for two particles
is reduced to three-dimensions  
by integrating out the time-component of relative
momentum~\cite{Phillips96}, the resulting equal-time (ET) formalism is quite similar 
to an instant form of relativistic quantum theory.
Restricting consideration to positive-energy particles, there is a
hamiltonian of the form $H = \epsilon_1 + \epsilon_2 + v$
in an arbitrary frame.  
When all contributions to $v$ from the reduction of a 
covariant four-dimensional
theory are kept, the hamiltonian in a frame where total momentum
is ${\bf P}$ has eigenvalue $E_{\bf P}$, 
corresponding to a fixed value of mass $M$.  
However when some contributions to the 
three-dimensional interaction are neglected, the eigenvalue 
fails to correspond to a fixed value of mass $M$.  This
motivates the study of sufficient conditions on the instant 
interaction to guarantee that the mass is fixed.   

In this paper, we show that use of an approximate boost generator
is sufficient to derive a simple boost rule for the
interaction, $v$, such that the instant two-body problem
has eigenvalue $E_{\bf P}$ corresponding to a fixed value of mass. 
The boost generator satisfies all but one of the required
commutator relations.  The analysis 
provides very simple and direct relationships of 
vertex functions, wave functions 
and t-matrices of the two-body problem in different frames.    
The results are applicable to calculations of 
matrix elements of an external current, for example form factor
calculations based upon the ET formalism, where   
the initial state or final state, or both, must have 
nonzero total momentum.

The basic requirement of Poincar\'e invariance is that states must
transform under a unitary representation of the Poincar\'e group.
The ten generators for translations in time, translations in space,
boosts and rotations are
the hamiltonian operator $H$, the operator for total momentum
${\bf P}$, the boost operator ${\bf K}$ and
the angular momentum operator ${\bf J}$.  
They obey well-known commutations relations 
(see Ref.~{\cite{Weinberg}, for example). When there is no interaction, 
all the required commutation relations may be satisfied by taking
each generator to be a sum of single particle generators. 
The interaction must be translationally and rotationally invariant
and must satisfy additional nontrivial constraints from the
commutation relations 
\begin{equation}
[K_i, P_j] = i H \delta_{ij}, 
\label{eq:[K,P]} 
\end{equation}
\begin{equation}
[{\bf K}, H] = i {\bf P}, 
\label{eq:[K,H]}  
\end{equation}
\begin{equation}
[K_i, K_j] = -i\epsilon_{ijk} J_k. 
\label{eq:[K,K]}
\end{equation}  
Standard rules of quantum mechanics apply, i.e., 
$[{\bf r}_j,{\bf p}_k] = i\delta_{jk}$,
where $\hbar = 1$.  

In the instant form of dynamics, total momentum and angular momentum
operators are, for spinless particles,
${\bf P} = {\bf p}_1 + {\bf p}_2$ and ${\bf J} = {\bf r}_1
\times {\bf p}_1 + {\bf r}_2 \times {\bf p}_2$.
For the case of two spinless particles with equal masses,
the hamiltonian for instant dynamics is
\begin{equation}
H =  \epsilon_1 + \epsilon_2 + v  = H_0 + v ,
\label{eq:H}
\end{equation}
where $\epsilon_i$ is the kinetic energy of
the ith particle,  
\begin{eqnarray}
\epsilon_1 = \epsilon_1({\bf p};{\bf P}) \equiv \sqrt{ m^2 + \left({1 \over 2}{\bf P} + {\bf p}\right)^{2} } ,
\nonumber \\
\epsilon_2 = \epsilon_2({\bf p};{\bf P}) \equiv \sqrt{ m^2 + \left({1 \over 2}{\bf P} - {\bf p}\right)^{2} }  .
\end{eqnarray}
and ${\bf p} = {1 \over 2} ({\bf p}_1 - {\bf p}_2)$.

Bakamjian and Thomas~\cite{Bakamjian53} derived the form of the boost operator. 
For a free boost, the operator is
\begin{equation}
{\bf K}_0 = {1 \over 2} \left( {\bf r}_1 \epsilon_1 + \epsilon_1 {\bf r}_1\right) 
+ {1 \over 2} \left( {\bf r}_2 \epsilon_2 + \epsilon_2 {\bf r}_2 \right).
\end{equation}
When the interaction $v$ is present in the hamiltonian, there is 
an interaction part of the boost operator that is given 
approximately by
\begin{equation}
{\bf K}_v = {1 \over 2} \left( {\bf R} v  + v {\bf R} \right),
\end{equation}
where ${\bf R} = {1 \over 2}({\bf r}_1 + {\bf r}_2)$.  
With these definitions, Eq.~(\ref{eq:[K,P]})  
is satisfied.  However, there is 
an error term of order $1/m^2$ in the commutator in Eq.~(\ref{eq:[K,K]}).
Thus, the boost generator ${\bf K}$  
is approximate.  Accepting this error, the
interaction $v$ must take an appropriate, 
but unknown, form consistent with Eq.~(\ref{eq:[K,H]}).    

Noting that the free-boost operator, ${\bf K}_0$, and free hamiltonian
$H_0$, obey
Eq.~(\ref{eq:[K,H]}), the interaction-dependent terms
in that commutation relation must sum to zero, i.e.,
\begin{equation}
[{\bf K}_0 , v] + [{\bf K}_v, H_0] + [{\bf K}_v, v] = 0 .
\end{equation}
This relation provides a constraint on the form of the interaction.
 It is equivalent to, after some algebraic manipulations,
\begin{eqnarray}
[{\bf R}, H_0] v + v [{\bf R}, H_0] +{1 \over 2} H [{\bf R}, v]
+ {1 \over 2} [{\bf R}, v] H  \nonumber \\
+ {1 \over 2} \Delta {\bf r} v - {1 \over 2} v {\bf r} \Delta  = 0,
\label{eq:comm}
\end{eqnarray}
where
\begin{equation}
\Delta \equiv \epsilon_1 - \epsilon_2 ,
\end{equation}
and ${\bf r} \equiv {\bf r}_1 - {\bf r}_2$.

Our goal is to determine $H$ such that its eigenvalue is
\begin{equation}
H |E_{{\bf P}}\rangle = E_{{\bf P}} |E_{{\bf P}}\rangle,
\end{equation}
which corresponds to an eigenstate of mass $M$.  We find that
it is sufficient to consider 
a diagonal matrix element of Eq.~(\ref{eq:comm})
in the state $|E_{{\bf P}}\rangle$, with $H$ replaced 
by its eigenvalue $E_{{\bf P}}$, as follows.
\begin{eqnarray}
\langle E_{{\bf P}}| \biggr( [{\bf R}, H_0] v + v [{\bf R}, H_0]
+ E_{{\bf P}} [{\bf R}, v] + \nonumber \\
{1 \over 2} \Delta {\bf r} v - {1 \over 2} v {\bf r} \Delta \biggr)
|E_{{\bf P}}\rangle = 0 .
\label{eq:CR1}
\end{eqnarray}
The $v^2$ terms in the third and fourth terms of Eq.~(\ref{eq:comm})
are eliminated in favor of the energy eigenvalue
$E_{{\bf P}}$.  

Some additional manipulations are needed to produce a solvable equation
for the interaction in which the boost velocity
${\bf P}/E_{{\bf P}}$ appears.  
In the last two terms involving $\Delta$,
$H/E_{{\bf P}}$ is inserted to the left of $\Delta {\bf r} v$ and to the right of
$v {\bf r}\Delta$.
Using $H_0 \Delta = 2 {\bf p} \cdot{\bf P} $, noting that
$[{\bf r}, \Delta ]= 2 [{\bf R}, H_0]$ and using
$v[{\bf R},H_0]v = {1 \over 2} (H - H_0)
[{\bf R},H_0]v + {1 \over 2} [{\bf R},H_0]v (H - H_0)$ and the
eigenvalue condition to replace $H$,
we arrive at the following equation.
\begin{eqnarray}
\langle E_{{\bf P}}|\biggr\{
\left( 1 + \frac{H_0}{ E_{{\bf P}}} \right)
\frac{[{\bf R}, H_0]}{2E_{{\bf P}}} v +
v \frac{[{\bf R}, H_0]}{2E_{{\bf P}}}
\left( 1 + \frac{H_0}{ E_{{\bf P}}} \right) \nonumber \\
+ [{\bf R}, v]  + \frac{ {\bf p} \cdot{\bf P} }{E_{{\bf P}}^2}{\bf r} v
- v {\bf r}  \frac{ {\bf p} \cdot{\bf P} }{E_{{\bf P}}^2} \biggr\}
|E_{{\bf P}}\rangle = 0  ,
\label{eq:CR3}
\end{eqnarray}
This equation is linear in $v$ and it may be solved in momentum space
 in order to 
determine the required form of the interaction.

Momentum space matrix elements involve
\begin{eqnarray}
\int \frac{d^3p'}{(2\pi)^3}\frac{d^3p}{(2\pi)^3}
\langle E_{{\bf P}} | {\bf p}'; {\bf P}\rangle
 {\cal B}({\bf p}',{\bf p};{\bf P})
\langle{\bf p};{\bf P} | E_{{\bf P}}\rangle = 0,
\label{eq:pspace}
\end{eqnarray}
where
\begin{equation}
 {\cal B}({\bf p}',{\bf p};{\bf P}) \equiv
\left[ {\bf C}({\bf p}'; {\bf P})  + {\bf C}({\bf p}; {\bf P})
+ {\bf D}_{op}\right] v({\bf p}',{\bf p};{\bf P})  ,
\end{equation}
\begin{equation}
{\bf C}({\bf p};{\bf P}) =
\left( 1 + \frac{d}{E_{{\bf P}}}  \right)
\frac{1}{2E_{{\bf P}}} \frac{\partial d}{\partial {\bf P}}
\end{equation}
with $d \equiv \epsilon_1({\bf p};{\bf P}) + \epsilon_2({\bf p};{\bf P}) $
being the value of $H_0$ and
\begin{equation}
{\bf D}_{op} \equiv
\frac{\partial }{\partial {\bf P}}
+ \frac{{\bf p}'\cdot{\bf P}}{E_{{\bf P}}^2}
\frac{\partial}{\partial{\bf p}'}
+ \frac{{\bf p}\cdot{\bf P}}{E_{{\bf P}}^2}
\frac{\partial}{\partial{\bf p}} .
\label{eq:Dop}
\end{equation}

It is sufficient that ${\cal B}({\bf p}',{\bf p};{\bf P}) = 0$ in Eq.~(\ref{eq:pspace})
in order to obtain a solution. 
The form of this partial differential equation for $v$ suggests that
the solution should have the form
\begin{equation}
v({\bf p}',{\bf p};{\bf P}) =
f({\bf p}';{\bf P}) \tilde{v}({\bf p}', {\bf p}; {\bf P})
f({\bf p};{\bf P}) ,
\label{eq:ff'vtilde}
\end{equation}
leading to
\begin{eqnarray}
\left( {\bf C}({\bf p};{\bf P}) +
\frac{1}{f({\bf p};{\bf P})} {\bf D}_{op}
f({\bf p};{\bf P})  \right) \nonumber \\
+  \left({\bf C}({\bf p}';{\bf P}) +
\frac{1}{f({\bf p}';{\bf P})} {\bf D}_{op}
f({\bf p}';{\bf P})  \right)
\nonumber \\
+ \frac{1}{\tilde{v}({\bf p}', {\bf p}; {\bf P})}
{\bf D}_{op} \tilde{v}({\bf p}', {\bf p}; {\bf P}) = 0.
\label{eq:sep}
\end{eqnarray}
Finally, there is a boundary condition:  when ${\bf P} = 0$,
$\tilde{v}$ must be the c.m. frame interaction that
determines the mass eigenvalue.
This requires that $f({\bf p};{\bf 0}) = 1$.

A solution of the equation may be obtained such that
${\bf D}_{op} \tilde{v} = 0$.  The form of $\tilde{v}$
may be deduced from this condition, which leads to the conclusion that
$\tilde{v}$ depends upon three rotational scalars: ${\bf p}_c^2$,
${\bf p}_c^{'2}$, and ${\bf p}_c\cdot {\bf p}_c'$, where
\begin{equation}
{\bf p}_c \equiv {\bf p} - \frac{({\bf p}\cdot {\bf P}) {\bf P}}
{E_{{\bf P}}( E_{{\bf P}} + M) } ,
\label{eq:pc}
\end{equation}
and
\begin{equation}
{\bf p}'_c \equiv {\bf p}' - \frac{({\bf p}'\cdot {\bf P}) {\bf P}}
{E_{{\bf P}}( E_{{\bf P}} + M) } .
\label{eq:pc'}
\end{equation}
Moreover, the boundary condition shows that
\begin{equation}
\tilde{v}({\bf p}',{\bf p};{\bf P}) = v_c({\bf p}_c',{\bf p}_c),
\label{eq:vtildevc}
\end{equation}
where $v_c$ is the c.m. frame interaction.
In the c.m. frame, ${\bf p}_c \rightarrow {\bf p}$ and 
${\bf p}'_c\rightarrow {\bf p}'$ are the standard
relative momenta.
If the total momentum is
in the $z$ direction, the z-component of relative momentum 
is simply $p_{cz} =  p_z/\gamma$, where $\gamma = E_{{\bf P}}/M$.
Thus, the relative momentum is Lorentz contracted along the 
direction of ${\bf P}$. 
The components of relative momenta perpendicular to the total
momentum are unaffected: ${\bf p}_{c\perp} = {\bf p}_{\perp}$.
The same rule applies to ${\bf p}_c'$.  It follows that ${\bf p}_c^2 =
{\bf p}^2 - ({\bf p}\cdot {\bf P})^2/E_{{\bf P}}^2$, and
similarly for ${\bf p}_c^{\prime 2}$.

A straightforward calculation shows that
\begin{eqnarray}
{\bf D}_{op} {\bf p}_c^2 = 0, ~~
{\bf D}_{op} {\bf p}_c^{'2} = 0, ~~
{\bf D}_{op} {\bf p}_c\cdot {\bf p}'_c = 0.
\label{eq:D_pc2}
\end{eqnarray}
It follows by the chain rule of differentiation that
\begin{equation}
{\bf D}_{op} v_c({\bf p}_c', {\bf p}_c) = 0.
\end{equation}
Thus, the term in Eq.~(\ref{eq:sep}) that involves $\tilde{v}$
vanishes identically when $\tilde{v}$ is identified as the c.m.
frame interaction evaluated with the relative momenta of
Eqs.~(\ref{eq:pc}) and (\ref{eq:pc'}).

It follows that the remaining two parts of
Eq.~(\ref{eq:sep}) that involve functions of ${\bf p}$ and
${\bf p}'$ must vanish separately.  Their sum equals zero
and they depend upon independent arguments.  Therefore,
the function $f({\bf p};{\bf P})$ must be a solution of the partial
differential equation in two variables
\begin{equation}
 {\bf C}({\bf p};{\bf P})
f({\bf p};{\bf P})+ {\bf D}_{op}
f({\bf p};{\bf P}) = 0.
\label{eq:PDEf}
\end{equation}

Instead of solving this equation directly, it is simpler to deduce
the solution from a consistency argument.
We consider the integral equation for a bound-state
vertex function in an arbitrary frame, and
the corresponding equation in the c.m. frame,
in order to deduce how $f$ should depend upon the momenta.

Using Eqs.~(\ref{eq:ff'vtilde}) and (\ref{eq:vtildevc}), 
the bound state vertex in an arbitrary frame
obeys the integral equation
\begin{eqnarray}
\Gamma ({\bf p}';{\bf P}) = \int \frac{d^3p}{(2\pi)^3}
f({\bf p}';{\bf P}) v_c({\bf p}_c', {\bf p}_c) f({\bf p}; {\bf P})
\nonumber \\
\frac{1}{E_{{\bf P}} - \epsilon_1({\bf p};{\bf P})
 - \epsilon_2({\bf p};{\bf P})} \Gamma({\bf p};{\bf P}).
\label{eq:Gamma}
\end{eqnarray}
By inspection, it is necessary that the solution should have the form
\begin{equation}
\Gamma ({\bf p}';{\bf P}) =
f({\bf p}';{\bf P}) \Gamma_c ({\bf p}'_c).
\label{eq:boostGamma}
\end{equation}
Moreover, $\Gamma_c$ should satisfy the corresponding 
equation in the c.m. frame,
\begin{equation}
\Gamma_c ({\bf p}'_c) = \int \frac{d^3p_c}{(2\pi)^3}
v_c({\bf p}_c', {\bf p}_c)
\frac{1}{ M - 2\epsilon_c({\bf p}_c)} \Gamma_c({\bf p}_c),
\label{eq:Gammac}
\end{equation}
where $\epsilon_c({\bf p}_c) \equiv \sqrt{m^2 + {\bf p}_c^2}$.
Noting that $d^3p = (E_{{\bf P}}/M) d^3p_c$, this will be true if
\begin{eqnarray}
f^2({\bf p};{\bf P}) = \frac{M}{E_{{\bf P}}} \left(\frac{ E_{{\bf P}}
- \epsilon_1({\bf p}; {\bf P}) - \epsilon_2({\bf p}; {\bf P})  }
{M - 2 \epsilon_c({\bf p}_c)} \right)
\nonumber \\
 = \frac{M}{E_{{\bf P}}} \left(\frac{M + 2\epsilon_c({\bf p}_c)}
{ E_{{\bf P}} + \epsilon_1({\bf p};{\bf P}) +\epsilon_2({\bf p};{\bf P})} \right)
\frac{1}{1 - \Delta^2/E_{{\bf P}}^2} .
\label{eq:fsol}
\end{eqnarray}
The second form follows after some algebra and it shows that there
is no possibility of a singularity because $|\Delta|/E_{{\bf P}} =
 |2 {\bf p}\cdot {\bf P|}/[(\epsilon_1+\epsilon_2)E_{{\bf P}}] <1 $.
For relative momenta $|{\bf p}| < m$ and $|{\bf p}'| < m$,
$|\Delta|/E_{{\bf P}} << 1$ and $f \approx M/E_{{\bf P}}$.

Consistency requires that the $f({\bf p};{\bf P})$ of Eq.~(\ref{eq:fsol}) should satisfy
the partial differential equation
(\ref{eq:PDEf}).  This may be verified straightforwardly by rewriting
(\ref{eq:PDEf}) in terms of $f^2$, and noting Eq.~(\ref{eq:D_pc2}).  It
completes the proof that
Eq.~(\ref{eq:CR3}) is satisfied exactly. 

For the two-body problem, the procedure noted above is sufficient
to provide the following interesting result.  
If an arbitrary, rotationally and translationally
invariant interaction in the c.m. frame
of the form $v_c({\bf p_c}', {\bf p}_c) $
defines the mass eigenvalue $M$ by solution of the c.m. frame equation,
\begin{equation}
[M - 2 \epsilon'_c({\bf p}_c') ]\Psi_c({\bf p}'_c) = \int \frac{d^3p_c}{(2\pi)^3}
v_c({\bf p}_c', {\bf p}_c) \Psi_c({\bf p}_c),
\end{equation}
then in another frame an instant-form equation corresponding 
to the same mass eigenvalue is
\begin{eqnarray}
[E_{{\bf P}} - \epsilon_1({\bf p}';{\bf P}) -
\epsilon_2({\bf p}';{\bf P})] \Psi({\bf p}';{\bf P}) =  \nonumber \\
\int \frac{d^3p}{(2\pi)^3}
v({\bf p}',{\bf p};{\bf P}) \Psi({\bf p};{\bf P})
\end{eqnarray}
where
\begin{equation}
v({\bf p}',{\bf p};{\bf P}) =
f({\bf p}';{\bf P}) v_c({\bf p}'_c, {\bf p}_c)
f({\bf p};{\bf P})
\label{eq:boostv}
\end{equation}
and the momenta ${\bf p}_c$ and ${\bf p}_c'$ in Eq.~(\ref{eq:boostv})
are defined in terms of total momentum ${\bf P}$ and 
relative momentum ${\bf p}$ in the arbitrary frame
as in Eqs.~(\ref{eq:pc}) and (\ref{eq:pc'}), while
$f({\bf p};{\bf P})$ is defined as in Eq.~(\ref{eq:fsol}).

Reference~\cite{CPS75} discusses an approximate boost rule that 
that is similar to the one given here.  If the mass eigenvalue $M$
and energy $E_{\bf P}$ are replaced by $2 \epsilon_c$ and 
$\epsilon_1 + \epsilon_2$, respectively, then the c.m. momentum
of Eq.~(\ref{eq:pc}) becomes the c.m. momentum ${\bf k}$ 
of Coester et al \cite{CPS75err}.  This is the no-interaction limit.  
The $f$ factors in the interaction are somewhat different from those of
Ref.~\cite{CPS75} in the same limit.   
The differences may be small
but the interaction of Ref.~\cite{CPS75} does not provide an exactly constant
mass eigenvalue.  

The vertex function in the c.m. frame, $\Gamma_c({\bf p}_c'$ of Eq.~(\ref{eq:Gammac}),
is related to the vertex function in an arbitrary frame,
$\Gamma({\bf p}';{\bf P})$, as in Eq.~(\ref{eq:boostGamma}).
The wave function in the c.m. frame,
$\Psi_c({\bf p}_c) = [M - 2 \epsilon_c({\bf p}_c)]^{-1}\Gamma_c({\bf p}_c)$
 and the wave function in an arbitrary frame,
$\Psi({\bf p};{\bf P}) = [E_{{\bf P}} - \epsilon_1({\bf p};{\bf P})
 - \epsilon_2({\bf p};{\bf P}) ]^{-1}
\Gamma({\bf p};{\bf P})$,
are related by
\begin{eqnarray}
\Psi({\bf p};{\bf P}) &=& f({\bf p};{\bf P}) \left(
\frac{M - 2 \epsilon_c({\bf p}_c)}{E_{{\bf P}} -
\epsilon_1({\bf p};{\bf P}) - \epsilon_2({\bf p};{\bf P}}\right)
\Psi_c({\bf p}_c),
\nonumber \\
 &=& \frac{M}{E_{\bf P} f({\bf p};{\bf P})}
\Psi_c({\bf p}_c).
\end{eqnarray}
The wave function in an arbitrary frame
is a factor times 
the c.m. frame wave function evaluated at
the appropriate arguments.  This is a simple form of 
unitary transformation that guarantees the preservation of
the normalization in the following sense, using $d^3p_c = d^3p M/E_{\bf P}$, 
\begin{equation}
\int \frac{d^3p}{(2\pi)^3} 
\left| \sqrt{\frac{E_{\bf P}}{M}} f({\bf p};{\bf P}) \Psi({\bf p};{\bf P}) \right|^2  = 
\int \frac{d^3p_c}{(2\pi)^3} | \Psi_c ({\bf p}_c)|^2. 
\end{equation}

Finally, the two-body t-matrix in the c.m. frame 
transforms in the same way as the interaction.
\begin{eqnarray}
t({\bf p}',{\bf p};{\bf P}) = f({\bf p}';{\bf P}) t_c({\bf p}_c',{\bf p}_c) 
f({\bf p};{\bf P}) ,
\label{eq:tmat}
\end{eqnarray}
where $t_c$ is the c.m. frame t-matrix.  This follows because
the integral in the Lippmann-Schwinger equation for the t-matrix
is the same as in Eq.~(\ref{eq:Gamma}).  When the boost relation 
of Eq.~(\ref{eq:tmat}) is used, with $f({\bf p};{\bf P})$ as in
Eq.~(\ref{eq:fsol}), one finds the Lippmann-Schwinger equation
for the t-matrix in the c.m. frame, which involves an 
integral as in Eq.~(\ref{eq:Gammac}).   

Although the boost rule is 
kinematical, it depends on the interaction through the eigenvalue.
It is applicable to form factor
calculations in the two body problem, for example, in the ET formalism. 
Inclusion of spin will be important to applications.  In the
ET formalism, the interaction involves a matrix element between 
plane-wave Dirac spinors, which should provide a useful starting point for
inclusion of spin degrees of freedom.  

In conclusion, an approximate, 
simple boost rule is determined that is sufficient to keep the mass eigenvalue
constant in the instant two-body problem for spinless particles.  Kinematical relations 
are given between interactions, vertex functions, wave functions 
and t-matrices in an arbitrary
frame and those in the c.m. frame.

\vspace{0.2in}
Support for this work from the U.S. Department of Energy is
gratefully acknowledged through
DOE grant DE-FG02-93ER-40762 at the University of Maryland and
DOE contract DE-AC05-84ER40150, under which the
Southeastern Universities Research Association (SURA) operates
the Thomas Jefferson National Accelerator Facility.
S. J. W. gratefully acknowledges support of SURA under its Sabbatical
Fellowship Program.

\end{document}